\documentclass[pra,aps,twocolumn,superscriptaddress,showpacs]{revtex4}
\usepackage{epsfig,amsmath}

\begin{document}

\title
{Quantum Nonlocality of N-qubit W states}

\author{Chunfeng Wu}
\affiliation{Department of Physics, National University of
Singapore, 2 Science Drive 3, Singapore 117542}
\author{Jing-Ling Chen}
\affiliation{Theoretical Physics Division, Nankai Institute of
Mathematics, Nankai University, Tianjin 300071, P. R. China}
\author{L. C. Kwek} \affiliation{Department of Physics,
National University of Singapore, 2 Science Drive 3, Singapore
117542} \affiliation{Nanyang Technological University, National
Institute of Education, 1, Nanyang Walk, Singapore 637616}
\author{C. H. Oh}
\email{phyohch@nus.edu.sg}\affiliation{Department of Physics,
National University of Singapore, 2 Science Drive 3, Singapore
117542}

\begin{abstract}
An experimental setup for testing quantum nonlocality of N qubits is
proposed. This method is a generalization of the optical setup
proposed by Banaszek and W\'odkiewicz \cite{KKe}. The quantum
nonlocality of N qubits can be obtained through its violation of
N-qubit Bell inequalities. The correlation function measured in the
experiment is described by the Wigner function. The effect of
inefficient detector is also considered.
\end{abstract}

 \maketitle

\section{Introduction}

Experimental verification on the conflict between quantum mechanics
and local realism for two particles have been well demonstrated in
several experiments \cite{Aspect81,2qubitex}. For a system with more
than two particles, for example three particles, experimental
verification of the violation of local realism is generally more
difficult. Recently, conflicts between quantum mechanics and local
realism for three qubits and four qubits have also been verified
\cite{ghztest,4qubitstest}. For N qubits, experimental observations
for violation of Bell inequalities are still lacking.

In Ref. \cite{KKe}, the authors proposed an optical setup for
testing quantum nonlocality in phase space for two qubits. The setup
essentially demonstrate quantum nonlocality based on phase space
measurement of the Wigner function using photon counting. The source
used in Ref. \cite{KKe} is a single photon incident on a 50:50 beam
splitter (see Figure \ref{sourcefig}). The generated state is
$|\psi(2)\rangle=\frac{1}{\sqrt{2}}(|10\rangle\pm |01\rangle)$
written in terms of the exit ports $a_1$ and $a_2$. For example,
$|10\rangle$ means one photon exits at port $a_1$ and no photon
exits at port $a_2$. $\pm$ can be realized by one phase shifter to
adjust the relative phase and the relative phase can be set to zero
without loss of generality. As pointed in Ref. \cite{KKe}, the
measuring apparatus consists of a beam splitter and photon counting
detector. The power transmission of the beam splitter is T. The
incidence of an excited coherent state $|\gamma\rangle$ on the
second input port of the beam splitter results in coherent
displacement. Namely, the action of the beam splitter is described
by $\hat{D}(\sqrt{1-T}\gamma)$ in the limit of $T\rightarrow 1$ and
$\gamma\rightarrow \infty$. The realistic measurement proposed in
Ref. \cite{KKe} is a test that the detectors are capable of
resolving the number of absorbed photons and $+1/-1$ are assigned
correspondingly to events in which an even/odd number of photons is
registered. In short, the correlation function measured is the
Wigner function. In the experiment, quantum nonlocality of two
qubits is revealed through violation of the CHSH inequality
\cite{CHSH69}. In this paper, we generalize the method to a
$N$-particle system.

For $N$ qubits, there have been umpteen Bell inequalities that have
been proposed \cite{Mermin90,Ardehali,BK,Werner01,ZB} to test
quantum nonlocality. The more general inequalities are the ones
given in \cite{Werner01,ZB}. These inequalities (\.{Z}B
inequalities) impose sufficient and necessary conditions for the
existence of local and realistic model. The \.{Z}B inequalities take
the form
\begin{eqnarray}
&& |\sum_{s_1,...s_N=\pm 1}S(s_1,...s_N)\times  \nonumber \\
&&\sum_{k_1,...k_N=1,2}s_1^{k_1-1}...s_n^{k_N-1}E(k_1,...k_N)|\leq
2^N, \label{ineqn}
\end{eqnarray}
where $E(k_1,...k_N)$ is correlation function measured in a
dichotomic-observable experiment. Specific choices give nontrivial
inequalities. For example, for $S(s_1,...,s_N)=\sqrt{2}(\cos
[-\frac{\pi}{4}+(s_1+...+s_N-N)\frac{\pi}{4}])$ one recovers the
MABK inequalities \cite{Mermin90, Ardehali, BK}.
 In the article, we propose a general experimental setup for
testing quantum nonlocality of $N$ qubits by using the \.{Z}B
inequalities. In section \ref{1}, we generalize the two-qubit
experimental scheme to $N$ qubits. We provide the general form of
the correlation functions measured in the scheme. In section
\ref{2}, we consider the cases for $N=3, 4$ as examples to show that
the experimental observation of quantum nonlocality can be linked to
the violation of 3-qubit and 4-qubit Bell inequalities. For three
qubits, the \.{Z}B inequality is not violated by all pure entangled
states \cite{Gisin91,ZB02}. In Refs. \cite{us,us2}, we constructed
Bell inequalities which are violated by any pure entangled state of
three qubits. Both 3-qubit \.{Z}B inequality and the Bell
inequalities given in Refs. \cite{us,us2} will be used to test
nonlocality in the proposed experiment. Effect of detector
inefficiency is also considered in the section.

\section{Proposed Experiment for Testing Quantum
Nonlocality}\label{1} 
\begin{figure}
\begin{center}
\epsfig{figure=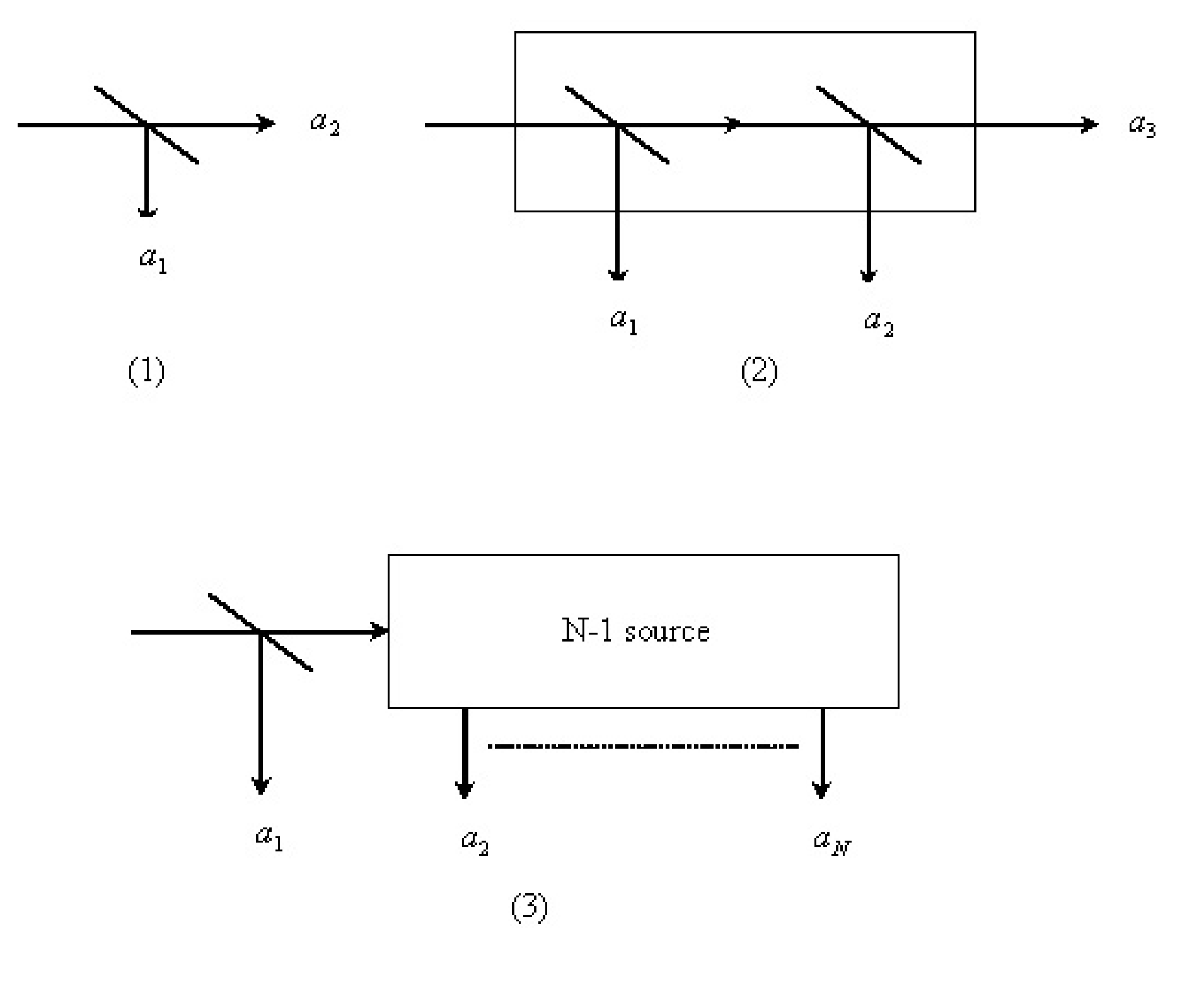,width=0.5\textwidth}
\end{center}
\caption{The sources proposed to generate N-qubit entangled states.
In (1), the beam splitter is a 50:50 one which generates a two-qubit
W state (a Bell state). In (2), the first beam splitter has a
reflectivity $R=\frac{1}{3}$ and the second beam splitter is a 50:50
one. These two beam splitters generate a three-qubit W state. We
have used a black box to represent the source. In (3), the beam
splitter has a reflectivity $R=\frac{1}{N}$. The box represents a
source which generates a (N-1)-qubit W state. The action of the beam
splitter and the (N-1) source is to generate a N-qubit W state.}
\label{sourcefig}
\end{figure}

We first consider the source for generating an
entangled state of $N$ qubits for the optical setup. For two qubits,
the source is a 50:50 beam splitter. The output state of the source
is $|\psi(2)\rangle=\frac{1}{\sqrt{2}}(|10\rangle+|01\rangle)$. For
three qubits, we look for a source that consists of a beam splitter
with reflectivity $R=\frac{1}{3}$ followed by a 50:50 beam splitter.
The output state is
$|\psi(3)\rangle=\frac{1}{\sqrt{3}}(|100\rangle+|010\rangle+|001\rangle)$.
Such a source is shown in Figure \ref{sourcefig} (2). In the Figure
\ref{sourcefig} (2), we use a black box to represent an effective
three-qubit source. For a four-qubit source, one can build on the
three-qubit source with an additional beam splitter. The beam
splitter in this case has a reflectivity $R=\frac{1}{4}$.
Generalizing, an $N$-qubit source for arbitrary $N$ qubit is
realized by a beam splitter with reflectivity $R=\frac{1}{N}$
followed by a $(N-1)$-qubit source. The overall action of
$(N-1)$-qubit source is that a single photon entering the box will
have equal chances of exiting at any of the $(N-1)$ ports. The
probability is $\frac{1}{N-1}$. The beam splitter has a
transmittance $T=\frac{N-1}{N}$. So in Figure \ref{sourcefig} (3),
the probability of a single photon exiting from each port
($a_2,...a_N$) of the box is
$\frac{1}{N-1}\frac{N-1}{N}=\frac{1}{N}$. The probability of a
single photon exiting from the reflected port $a_1$ of the beam
splitter is also $\frac{1}{N}$ since its reflectivity
$R=\frac{1}{N}$. Therefore the effective source generates an
$N$-qubit W state
$|\psi(N)\rangle=\frac{1}{\sqrt{N}}(|10...0\rangle+|01...0\rangle+...+|0...01\rangle)$.

\begin{figure}
\begin{center}
\epsfig{figure=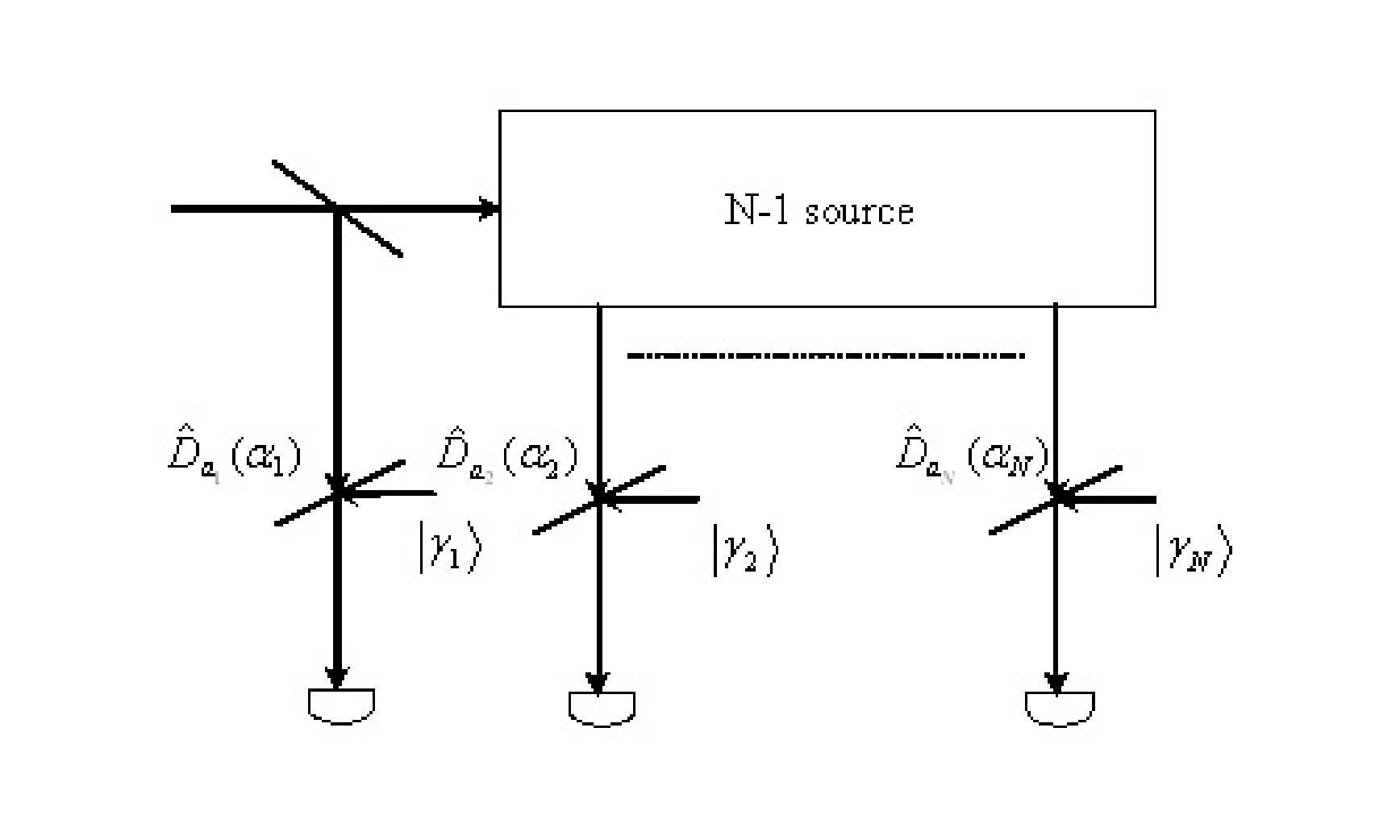,width=0.5\textwidth}
\end{center}
\caption{The optical setup for testing nonlocality of N qubits. The
beam splitter in the top line, which has a reflectivity $R=1/N$, and
the (N-1) source generate a N-qubit entangled state. Coherent
displacements $\hat{D}_{a_i}(\alpha_i)$ are realized by strong
coherent states $|\gamma_i\rangle$ ($\gamma_i\rightarrow\infty$)
injected into beam splitters which have transmittance close to 1
($T_i\rightarrow 1$). N detectors are put in the lowest line.}
\label{setupfig}
\end{figure}
The experimental setup is schematically shown in Figure
\ref{setupfig}. A single photon enters the source and exits with an
equal chance at each outgoing mode $a_i$. The measuring devices are
photon detectors preceded by beam splitters which have transmittance
$T_i\rightarrow 1$. Strong coherent states $|\gamma_i\rangle$ are
injected into the beam splitters. As a result, the beam splitters
give coherent displacements $\hat{D}_{a_i}(\alpha_i)$ with
$\alpha_i=\sqrt{1-T_i}\gamma_i$ on the input modes when
$\gamma_i\rightarrow \infty$ \cite{KKe}. By assigning $+1/-1$ to the
events that an even/odd number of photons has been registered, the
correlation function measured in the experiment is given by
\begin{eqnarray}
&&\Pi_{a_1a_2...a_N}(\alpha_1,\alpha_2,...,\alpha_N)= \nonumber \\
&&\langle\psi(N)|\hat{\Pi}_{a_1}(\alpha_1)\otimes\hat{\Pi}_{a_2}(\alpha_2)\otimes
...\otimes\hat{\Pi}_{a_N}(\alpha_N)|\psi(N)\rangle, \nonumber \\
\end{eqnarray}
where $\alpha_i$ is coherent displacement for the mode $a_i$.
$\hat{\Pi}_{a_i}(\alpha_i)$ is an operator defined as \cite{KKe}
\begin{eqnarray}
\hat{\Pi}_{a_i}(\alpha_i)&=&\hat{D}_{a_i}(\alpha_i)
\sum_{k=0}^{\infty}|2k\rangle\langle2k|\hat{D}_{a_i}^{\dagger}(\alpha_i)
\nonumber \\
&&-\hat{D}_{a_i}(\alpha_i)\sum_{k=0}^{\infty}|2k+1\rangle\langle2k+1|\hat{D}_{a_i}^{\dagger}(\alpha_i).
\end{eqnarray}
The operator $\hat{\Pi}_{a_i}(\alpha_i)$ is also of the form
\cite{KKe}
\begin{eqnarray}
\hat{\Pi}_{a_i}(\alpha_i)=\hat{D}_{a_i}(\alpha_i)(-1)^{\hat{n}_{a_i}}\hat{D}^{\dagger}_{a_i}(\alpha_i),
\end{eqnarray}
which is the displaced parity operator. Because that
\begin{eqnarray}
\hat{D}_{a_i}^{\dagger}(\alpha_i)|0\rangle&=&e^{-\frac{1}{2}|\alpha_i|^2}\sum_{m}(-\alpha_i)^m\frac{1}{\sqrt{m!}}|m\rangle  \nonumber \\
\hat{D}_{a_i}^{\dagger}(\alpha_i)|1\rangle&=&e^{-\frac{1}{2}|\alpha_i|^2}\{\sum_{m}(-\alpha_i)^m\alpha_i^*\frac{1}{\sqrt{m!}}|m\rangle
\nonumber \\
&&+\sum_{m}(-\alpha_i)^m\frac{\sqrt{m+1}}{\sqrt{m!}}|m+1\rangle\}
\end{eqnarray}
for the state $|\Psi(N)\rangle$, we have
\begin{eqnarray}
&&\Pi_{a_1a_2...a_N}(\alpha_1,\alpha_2,...,\alpha_N)= \nonumber \\
&&\frac{1}{N}(4|\sum_{i=1}^{N}\alpha_i|^2-N)e^{-2\sum_{i=1}^{N}|\alpha_i|^2}.
\end{eqnarray}
For $N$ qubits, by measuring correlation functions and using \.{Z}B
inequalities, one can hopefully show violation of $N$-qubit Bell
inequality. In the next section, we investigate quantum nonlocality
of 3 and 4 qubits.

\section{Examples}\label{2}
In this section, we generalize the scheme to 3 qubits first. For
three qubits, the source of nonclassical radiation is a single
photon incident on a beam splitter with transmittance $T=2/3$ and
reflectivity $R=1/3$ followed by a 50:50 beam splitter, which
generates a three-qubit W state. The measuring devices are photon
counters with beam splitters placed in front of them. The beam
splitters have the transmission coefficient close to one with strong
coherent beams injected into the auxiliary ports. In this limit, the
beam splitters effectively perform coherent displacements
$\hat{D}_{a_1}(\alpha_1)$, $\hat{D}_{a_2}(\alpha_2)$ and
$\hat{D}_{a_3}(\alpha_3)$ on the three ports (modes) of the input
field. The correlation function measured is
\begin{eqnarray}
&&\Pi_{a_1a_2a_3}(\alpha_1,\alpha_2,\alpha_3) \nonumber \\
&&=\langle\psi(3)|\hat{\Pi}_{a_1}(\alpha_1)\otimes\hat{\Pi}_{a_2}(\alpha_2)\otimes\hat{\Pi}_{a_3}(\alpha_3)|\psi(3)\rangle
\nonumber \\
&&=\frac{1}{3}e^{-2(|\alpha_1|^2+|\alpha_2|^2+|\alpha_3|^2)}(4|\alpha_1+\alpha_2+\alpha_3|^2-3).
\end{eqnarray}
We next construct a Bell quantity from the 3-qubit \.{Z}B
inequality:
\begin{eqnarray}
{\cal B}_{\rm 3qubits}&=&\Pi_{a_1a_2a_3}(\alpha^1_1,\alpha^1_2,\alpha^2_3)+\Pi_{a_1a_2a_3}(\alpha^1_1,\alpha^2_2,\alpha^1_3) \nonumber \\
&+&\Pi_{a_1a_2a_3}(\alpha^2_1,\alpha^1_2,\alpha^1_3)-\Pi_{a_1a_2a_3}(\alpha^2_1,\alpha^2_2,\alpha^2_3).
\nonumber \\
\end{eqnarray}
where $\alpha^i_j$ with $i=1,2$ and $j=1,2,3$ are the two
experimental settings of the coherent displacements for ports $a_1,
a_2$ and $a_3$ respectively. For a local realistic theory, ${\cal
B}_{\rm 3qubits}\le 2$. Unfortunately 3-qubit \.{Z}B inequality does
not reveal quantum nonlocality since a numerical calculation gives
${\cal B}_{\rm 3qubits} \leq 2$. A possible reason is that the
degree of quantum nonlocality depends not only on the given
entangled state but also on the Bell operator \cite{Bellop}. Hence
the result means that 3-qubit \.{Z}B inequality may not reveal
quantum nonlocality for displacement measurements on the particles.

Recently new Bell inequalities for three qubits have been proposed
in Refs. \cite{us,us2}. The Bell inequalities are violated for any
pure entangled state. We next show that quantum nonlocality of three
qubits can be exhibited in the proposed experimental scheme using
the correlation-form inequality given in \cite{us2}.

Unlike the 3-qubit \.{Z}B inequality where there are only
three-particle correlation functions, the recent Bell inequality for
three qubits contains two-particle correlation functions and
one-particle correlation functions \cite{us2}. The two-particle
correlation functions and one-particle correlation functions are
given by
\begin{eqnarray}
&&\Pi_{a_1a_2}(\alpha_1,\alpha_2)=\langle\psi(3)|\hat{\Pi}_{a_1}(\alpha_1)\otimes\hat{\Pi}_{a_2}(\alpha_2)\otimes{\bf
1}|\psi(3)\rangle,
\nonumber \\
&&\Pi_{a_1a_3}(\alpha_1,\alpha_3)=\langle\psi(3)|\hat{\Pi}_{a_1}(\alpha_1)\otimes{\bf
1}\otimes\hat{\Pi}_{a_3}(\alpha_3)|\psi(3)\rangle,
\nonumber \\
&&\Pi_{a_2a_3}(\alpha_2,\alpha_3)=\langle\psi(3)|{\bf
1}\otimes\hat{\Pi}_{a_2}(\alpha_2)\otimes\hat{\Pi}_{a_3}(\alpha_3)|\psi(3)\rangle,
\nonumber \\
&&\Pi_{a_1}(\alpha_1)=\langle\psi(3)|\hat{\Pi}_{a_1}(\alpha_1)\otimes{\bf
1}\otimes{\bf 1}|\psi(3)\rangle,
\nonumber \\
&&\Pi_{a_2}(\alpha_2)=\langle\psi(3)|{\bf
1}\otimes\hat{\Pi}_{a_2}(\alpha_2)\otimes{\bf 1}|\psi(3)\rangle,
\nonumber \\
&&\Pi_{a_3}(\alpha_3)=\langle\psi(3)|{\bf 1}\otimes{\bf
1}\otimes\hat{\Pi}_{a_3}(\alpha_3)|\psi(3)\rangle.
\nonumber \\
\end{eqnarray}
The two-particle correlation functions are measured when one of
three observers does not perform any measurement on his detector.
The one-particle correlation functions are measured when two of
three observers do not perform any measurement on their detectors.
By similar calculations to the one for three-particle correlation
function, we have
\begin{eqnarray}
&&\Pi_{a_ia_j}(\alpha_i,\alpha_j)=\frac{1}{3}e^{-2(|\alpha_i|^2+|\alpha_j|^2)}(4|\alpha_i+\alpha_j|^2-1),
\nonumber \\
&&\Pi_{a_i}(\alpha_i)=\frac{1}{3}e^{-2(|\alpha_i|^2)}(4|\alpha_i|^2+1),
\nonumber \\
\end{eqnarray}
where $i,j=1,2,3$. Using a correlation form of the inequalities for
three qubits given in Ref. \cite{us2}, we construct a new Bell
quantity for three qubits
\begin{eqnarray}
{\cal B}'_{\rm 3qubits}&=&-\Pi_{a_1a_2a_3}(\alpha^1_1,\alpha^1_2,\alpha^1_3)+\Pi_{a_1a_2a_3}(\alpha^1_1,\alpha^1_2,\alpha^2_3) \nonumber \\
&+&\Pi_{a_1a_2a_3}(\alpha^1_1,\alpha^2_2,\alpha^1_3)+\Pi_{a_1a_2a_3}(\alpha^2_1,\alpha^1_2,\alpha^1_3) \nonumber \\
&-&\Pi_{a_1a_2a_3}(\alpha^2_1,\alpha^2_2,\alpha^2_3)-\Pi_{a_1a_2}(\alpha^1_1,\alpha^2_2)\nonumber \\
&-&\Pi_{a_1a_2}(\alpha^2_1,\alpha^1_2)-\Pi_{a_1a_2}(\alpha^2_1,\alpha^2_2) \nonumber \\
&-&\Pi_{a_1a_3}(\alpha^1_1,\alpha^2_3)-\Pi_{a_1a_3}(\alpha^2_1,\alpha^1_3) \nonumber \\
&-&\Pi_{a_1a_3}(\alpha^2_1,\alpha^2_3)-\Pi_{a_2a_3}(\alpha^1_2,\alpha^2_3) \nonumber \\
&-&\Pi_{a_2a_3}(\alpha^2_2,\alpha^1_3)-\Pi_{a_2a_3}(\alpha^2_2,\alpha^2_3)\nonumber
\\
&+&\Pi_{a_1}(\alpha^1_1)+\Pi_{a_2}(\alpha^1_2)+\Pi_{a_3}(\alpha^1_3).
\end{eqnarray}
Local realism theories impose the upper bound value of 3 for the
Bell quantity ${\cal B}'_{\rm 3qubits}$. By taking the coherent
displacements as $\alpha^1_1=\alpha^1_2=\alpha^1_3=0.471669$,
$\alpha^2_1=\alpha^2_2=\alpha^2_3=-0.0205849$, ${\cal B}'_{\rm
3qubits}=3.1605$ which is greater than 3. Thus one can detect
quantum nonlocality of a three-qubit system in the proposed
experiment.

For four qubits, the effective 4-qubit source generates a 4-qubit W
state which takes the form
\begin{eqnarray}
|\psi(4)\rangle=\frac{1}{2}(|1000\rangle+|0100\rangle+|0010\rangle+|0001\rangle).
\end{eqnarray}
The correlation function measured is given by
\begin{widetext}
\begin{eqnarray}
\Pi_{a_1a_2a_3a_4}(\alpha_1,\alpha_2,\alpha_3,\alpha_4)
&=&\langle\psi(4)|\hat{\Pi}_{a_1}(\alpha_1)\otimes\hat{\Pi}_{a_2}(\alpha_2)\otimes\hat{\Pi}_{a_3}(\alpha_3)\otimes\hat{\Pi}_{a_4}(\alpha_4)|\psi(4)\rangle
\nonumber \\
&&=\frac{1}{4}e^{-2(|\alpha_1|^2+|\alpha_2|^2+|\alpha_3|^2+|\alpha_4|^2)}(4|\alpha_1+\alpha_2+\alpha_3+\alpha_4|^2-4).
\end{eqnarray}
\end{widetext}
Based on the 4-qubit \.{Z}B inequality, the Bell quantity ${\cal
B}_{\rm 4qubits}$ is constructed as
\begin{widetext}
\begin{eqnarray}
{\cal B}_{\rm
4qubits}&=&\Pi_{a_1a_2a_3a_4}(\alpha^1_1,\alpha^1_2,\alpha^1_3,\alpha^1_4)-\Pi_{a_1a_2a_3a_4}(\alpha^1_1,\alpha^1_2,\alpha^1_3,\alpha^2_4)
-\Pi_{a_1a_2a_3a_4}(\alpha^1_1,\alpha^1_2,\alpha^2_3,\alpha^1_4)-\Pi_{a_1a_2a_3a_4}(\alpha^1_1,\alpha^1_2,\alpha^2_3,\alpha^2_4)
\nonumber
\\
&-&\Pi_{a_1a_2a_3a_4}(\alpha^1_1,\alpha^2_2,\alpha^1_3,\alpha^1_4)-\Pi_{a_1a_2a_3a_4}(\alpha^1_1,\alpha^2_2,\alpha^1_3,\alpha^2_4)
-\Pi_{a_1a_2a_3a_4}(\alpha^1_1,\alpha^2_2,\alpha^2_3,\alpha^1_4)+\Pi_{a_1a_2a_3a_4}(\alpha^1_1,\alpha^2_2,\alpha^2_3,\alpha^2_4)
\nonumber \\
&-&\Pi_{a_1a_2a_3a_4}(\alpha^2_1,\alpha^1_2,\alpha^1_3,\alpha^1_4)-\Pi_{a_1a_2a_3a_4}(\alpha^2_1,\alpha^1_2,\alpha^1_3,\alpha^2_4)
-\Pi_{a_1a_2a_3a_4}(\alpha^2_1,\alpha^1_2,\alpha^2_3,\alpha^1_4)+\Pi_{a_1a_2a_3a_4}(\alpha^2_1,\alpha^1_2,\alpha^2_3,\alpha^2_4)
\nonumber \\
&-&\Pi_{a_1a_2a_3a_4}(\alpha^2_1,\alpha^2_2,\alpha^1_3,\alpha^1_4)+\Pi_{a_1a_2a_3a_4}(\alpha^2_1,\alpha^2_2,\alpha^1_3,\alpha^2_4)
+\Pi_{a_1a_2a_3a_4}(\alpha^2_1,\alpha^2_2,\alpha^2_3,\alpha^1_4)+\Pi_{a_1a_2a_3a_4}(\alpha^2_1,\alpha^2_2,\alpha^2_3,\alpha^2_4),
\nonumber \\
\end{eqnarray}
\end{widetext}
where $\alpha^i_j$ with $i=1,2$ and $j=1,2,3,4$ are two experimental
settings of the coherent displacements for mode $a_1, a_2, a_3$ and
$a_4$ respectively. By setting the experimental values of the
coherent displacement as
$\alpha^1_1=\alpha^1_2=\alpha^1_3=\alpha^1_4=-0.104749$,
$\alpha^2_1=\alpha^2_2=\alpha^2_3=\alpha^2_4=0.294117$, the maximum
value of ${\cal B}_{\rm 4qubits}$ is calculated numerically to be
5.14529. Since local realistic theories require ${\cal
B}_{4qubits}\le 4$, the violation of local realism is simply shown
from \.{Z}B inequality.

It should be mentioned that other factors should be taken into
account, such as detector inefficiencies, in practice. If one
considers the inefficiency of detector, one has to modify the
correlation slightly to account for the imperfections, which are
characterized by the quantum efficiency of detectors $\eta$ ($0\leq
\eta \leq 1$). For ideal detectors, $\eta=1$ and the correlation is
perfect. For non-ideal detectors, the correlation is modified by
$\Pi'_{a_1...a_N}(\alpha_1...\alpha_N)$. If we assume that all the
photon detectors have the same efficiencies $\eta$,
\begin{eqnarray}
&&\Pi'_{a_1a_2...a_N}(\alpha_1,\alpha_2,...,\alpha_N)= \nonumber \\
&&\langle\psi(N)|\hat{\Pi}'_{a_1}(\alpha_1)\otimes\hat{\Pi}'_{a_2}(\alpha_2)\otimes
...\otimes\hat{\Pi}'_{a_N}(\alpha_N)|\psi(N)\rangle, \nonumber \\
\end{eqnarray}
where $\hat{\Pi}'_{a_i}(\alpha_i)$ is an operator defined as
\begin{eqnarray}
\hat{\Pi}'_{a_i}(\alpha_i)&=&\hat{D}_{a_i}(\alpha_i)(-1)^{\eta\hat{n}_{a_i}}\hat{D}^{\dagger}_{a_i}(\alpha_i)
\nonumber \\
&=&\hat{D}_{a_i}(\alpha_i)(1-2\eta)^{\hat{n}_{a_i}}\hat{D}^{\dagger}_{a_i}(\alpha_i).
\end{eqnarray}
A straightforward calculation yields the modified correlation
function for the state $\psi(N)\rangle$
\begin{eqnarray}
&&\Pi'_{a_1a_2...a_N}(\alpha_1,\alpha_2,...,\alpha_N)= \nonumber \\
&&\frac{1}{N}\{(-2\eta)^2|\sum_{i=1}^{N}\alpha_i|^2+N(1-2\eta)\}e^{-2\eta\sum_{i=1}^{N}|\alpha_i|^2}.
\end{eqnarray}
The detector efficiency directly affects the experimental results.
As an example, we consider the 3-qubit experiment. Using the
modified correlation functions
\begin{eqnarray}
&&\Pi'_{a_1a_2a_3}(\alpha_1,\alpha_2,\alpha_3)= \nonumber \\
&&\frac{1}{3}\{(-2\eta)^2|\sum_{i=1}^{3}\alpha_i|^2+3(1-2\eta)\}e^{-2\eta\sum_{i=1}^{3}|\alpha_i|^2}
\nonumber \\
&&\Pi'_{a_ia_j}(\alpha_i,\alpha_j)= \nonumber \\
&&\frac{1}{3}\{(-2\eta)^2(|\alpha_i+\alpha_j|^2)+3-4\eta\}e^{-2\eta(|\alpha_i|^2+|\alpha_j|^2)}
\nonumber \\
&&\Pi'_{a_i}(\alpha_i)= \nonumber \\
&&\frac{1}{3}\{(-2\eta)^2(|\alpha_i|^2)+3-2\eta\}e^{-2\eta(|\alpha_i|^2)}
\end{eqnarray}
where $i,j=1,2,3$, a Bell quantity ${\cal B}^{\eta}_{\rm 3qubits}$
is constructed from the Bell inequality given in Ref. \cite{us2},
\begin{eqnarray}
{\cal B}^{\eta}_{\rm 3qubits}&=&-\Pi'_{a_1a_2a_3}(\alpha^1_1,\alpha^1_2,\alpha^1_3)+\Pi'_{a_1a_2a_3}(\alpha^1_1,\alpha^1_2,\alpha^2_3) \nonumber \\
&+&\Pi'_{a_1a_2a_3}(\alpha^1_1,\alpha^2_2,\alpha^1_3)+\Pi'_{a_1a_2a_3}(\alpha^2_1,\alpha^1_2,\alpha^1_3) \nonumber \\
&-&\Pi'_{a_1a_2a_3}(\alpha^2_1,\alpha^2_2,\alpha^2_3)-\Pi'_{a_1a_2}(\alpha^1_1,\alpha^2_2)\nonumber \\
&-&\Pi'_{a_1a_2}(\alpha^2_1,\alpha^1_2)-\Pi'_{a_1a_2}(\alpha^2_1,\alpha^2_2) \nonumber \\
&-&\Pi'_{a_1a_3}(\alpha^1_1,\alpha^2_3)-\Pi'_{a_1a_3}(\alpha^2_1,\alpha^1_3) \nonumber \\
&-&\Pi'_{a_1a_3}(\alpha^2_1,\alpha^2_3)-\Pi'_{a_2a_3}(\alpha^1_2,\alpha^2_3) \nonumber \\
&-&\Pi'_{a_2a_3}(\alpha^2_2,\alpha^1_3)-\Pi'_{a_2a_3}(\alpha^2_2,\alpha^2_3) \nonumber \\
&+&\Pi'_{a_1}(\alpha^1_1)+\Pi'_{a_2}(\alpha^1_2)+\Pi'_{a_3}(\alpha^1_3).
\end{eqnarray}
Since, for a local realistic description, ${\cal B}^{\eta}_{\rm
3qubits}\le 3$  and for quantum nonlocality, the Bell quantity
${\cal B}^{\eta}_{\rm 3qubits}$ has a maximum value of 3.1605, which
is greater than 3 when $\eta> 0.9804$. Thus, the nonlocality of a
three-qubit system exhibits in the proposed experiment if $\eta>
0.9804$. Quantum nonlocality of N qubits ( N is an arbitrary number)
can be tested in the proposed experiment in a similar way.

In the article, we propose an experimental setup for testing quantum
nonlocality of $N$ qubits by using $N$-qubit Bell inequalities. The
correlation function measured in the scheme is described by the
Wigner function. We also briefly consider the case in which the
detectors are inefficient and provide a threshold value above which
nonlocality observation can be obtained.

This work is supported by NUS academic research Grant No. WBS:
R-144-000-123-112.

\end{document}